\documentclass[12pt,technote,twoside]{IEEEtran}
\usepackage{latexsym,amssymb}
\usepackage{supertabular}

\newcommand{\C}[0]{{\mathbf{C}}}

\newcommand{\Z}{{\mathbf{Z}}}
\newcommand{\mat}[4]{\left(\begin{array}{rr}#1 & #2\\ #3 & #4 
\end{array}\right)
}

\newcommand{\nix}[1]{}

\def\qed{\quad{$\Box$}}

\def\trace{\mathop{{\rm tr}}\nolimits}
\def\ker{\mathop{{\rm ker}}\nolimits}
\def\image{\mathop{{\rm im}}\nolimits}

\def\supp{\mathop{{\rm supp}}\nolimits}
\def\Irr{\mathop{{\rm Irr}}\nolimits}
\def\IrrP(#1){\mathop{{\rm Irr}}\nolimits(\phi\,|\,#1)}

\newtheorem{theorem}{Theorem}

\newtheorem{lemma}[theorem]{Lemma}

\newcommand{\projrho}{\rho}
\newcommand{\ordrho}{\varrho}

\begin{document}

\title{Beyond Stabilizer Codes II: Clifford Codes}
\author{%
Andreas Klappenecker,
\thanks{A.K.~thanks the Santa Fe Institute for support through their
Fellow-at-Large program, and the European Community for support
through the grant IST-1999-10596 (Q-ACTA). M.R.~thanks the DFG for
support through the Graduiertenkolleg GRK-209/3-98.  Part of this work
has been done at the Tacheles, Berlin, during a concert of Remember
the Day.
}
\thanks{A.~Klappenecker is with the Department of Computer Science, 
Texas A\&M University, College Station, TX 77843-3112, USA (e-mail: klappi@cs.tamu.edu).}%
Martin R\"otteler 
\thanks{M.~R\"otteler is with the Institut f\"ur Algorithmen und
Kognitive Systeme, Forschungsgruppe Quantum Computing
(Professor Thomas Beth), Universit\"at Karlsruhe, Am Fasanengarten 5,
D-76\,128 Karlsruhe, Germany (e-mail: roettele@cs.tamu.edu).}%
}

\maketitle

\def\baselinestretch{1.66}
\small\normalsize

\begin{abstract}
\noindent Knill introduced a generalization of stabilizer codes, in
this note called Clifford codes. It remained unclear whether or not
Clifford codes can be superior to stabilizer codes.  We show that
Clifford codes \textit{are} stabilizer codes provided that the
abstract error group has an abelian index group.  In particular, if
the errors are modelled by tensor products of Pauli matrices, then the
associated Clifford codes are necessarily stabilizer codes.
\end{abstract}


\section{Introduction}
Quantum error control codes allow to protect the computational states
of a quantum computer against decoherence errors.  Almost all quantum
codes known today have been constructed as stabilizer codes,
cf.~\cite{ashikhmin00,calderbank98,gottesman96,rains99}.  Allowing the
protection of quantum systems of arbitrary finite dimension, we are led
to modify the notion of a stabilizer code in the following way:

Let $\ordrho\colon E\rightarrow {\mathcal U}(n)$ be a faithful unitary
irreducible ordinary representation of an abstract error group
$E$~\cite{klappeneckerPartI}. A {\em stabilizer code}\/ is defined to
be the joint eigenspace $Q$ of the representing matrices $\ordrho(n)$
for all $n\in N$, where $N$ is a normal subgroup of~$E$.  If $Q$ is
nontrivial, then $N$ is necessarily an {\em abelian} normal subgroup
of $E$.

We recover the definition of binary stabilizer codes as used
in~\cite{calderbank98,gottesman96} by taking~$E$ to be the generalized
extraspecial 2-group which is generated by $k$-fold tensor products of
Pauli matrices. The stabilizer codes derived within this error model have
been studied in great detail, see~\cite{calderbank98,gottesman96}.

The definition of stabilizer codes forces the normal subgroup to be
abelian.  A more general class of quantum error correcting codes -- in
this note called Clifford codes~-- has been introduced by Knill
in~\cite{knill96b}.  Clifford codes are derived with the help of
normal subgroups which are not necessarily abelian.

It remained unclear, whether or not it is possible to construct
Clifford codes that are better than stabilizer codes. We found
surprisingly good Clifford codes by computer search. However, we were
never able to beat the stabilizer codes. The main result of this note
partly explains this phenomenon: we find that each Clifford code is
actually a stabilizer code given that the error group has an abelian
index group. Therefore, the error models discussed in
\cite{ashikhmin00,calderbank98,gottesman96} cannot lead to Clifford
codes which are not stabilizer codes.

\section{Clifford Codes}
We will construct a quantum code $Q$ from a normal subgroup~$N$ of an
abstract error group~$E$.  The main properties of such a code~$Q$ are
determined by applying results from Clifford theory, hence the name
Clifford code.  The relevant results from Clifford theory can be found
in Huppert~\cite[Chapter 5]{huppert83} or any other standard text on
representation theory of finite groups.

Let $E$ be an abstract error group. Recall that the group $E$ has a
faithful irreducible ordinary representation $\ordrho\colon
E\rightarrow {\mathcal{U}}(n)$ of large degree $ \deg \ordrho =
(E\!:\! Z(E))^{1/2}.$ The errors are expressed as linear combinations
of the unitary $n\times n$ matrices $\ordrho(g)$ representing elements $g$ of the abstract error group $E$. 

The action of the representation $\ordrho$ on $\C^n$ induces an
irreducible $\C E$-module structure on the ambient space $\C^n$. 
Let $N$ be a normal subgroup of $E$, denoted by $N\trianglelefteq E$. 
 If
we view the ambient space $\C^n$ as a $\C N$-module, then we obtain a
decomposition into irreducible $\C N$-modules $gW$ of the form
$$ \C^n \cong \bigoplus_{i=1}^m \left\{\bigoplus_{g\in R} gW \right\},$$ where $R$ is
a transversal of the inertia group $T(W)$ in $E$, and $m$ is the
multiplicity of the module $gW$ in this decomposition. Recall
that the inertia group is defined by 
$$T(W) = \{ g\in E\,|\, gW\cong
W\,\}.$$ We define a quantum code $Q$ to be a homogeneous component
$$Q\cong\underbrace{W\oplus \cdots \oplus W}_{\textup{\footnotesize $m$-times}}$$
of this decomposition. Thus, $Q$ is a subspace of $\C^n$ which is also
endowed with the structure of a 
$\C N$-module. We call any quantum code $Q$ that can be
obtained by such a construction a {\em Clifford code}.

We need to introduce some more notation before we can discuss the error
correcting properties of a Clifford code $Q$. We define $Z(W)$ to be
the set of elements that act on $Q$ by scalar multiplication
$$Z(W)=\{ g\in T(W)\,|\, \exists \lambda \in \C\,\forall v\in Q:
gv=\lambda v\}.$$ 
The error correcting properties of the code
$Q$ are summarized by the following theorem. Although this theorem is
essentially contained in~\cite{knill96b}, we include it here to make this note self-contained:
\begin{theorem}
We keep the notation introduced above.
Let $\chi$ be the character of $N$ afforded by $W$.
Then  
$$ e_\chi=\frac{\chi(1)}{|N|} \sum_{n\in N} \chi(n^{-1})\ordrho(n)$$ is
an orthogonal projector onto $Q$. 
The code $Q$ is able to correct a
set of errors $\Sigma\subset E$ precisely when the condition $e^{-1}_1
e_2\not\in T(W)-Z(W)$ holds for all $e_1,e_2\in \Sigma$. The dimension of $Q$ is $m\chi(1)$. 
\end{theorem}
\proof We divide the proof into several steps.
\begin{enumerate}
\item[\textit{Step 1.}]  The matrix group $\ordrho(N)$ is isomorphic
to the abstract group~$N$, since $\ordrho$ is a faithful
representation. Since $\chi$ is an irreducible character of $N$, it
follows that $e_\chi$ is an idempotent in the group algebra
$\C[\ordrho(N)]\cong \C N$, cf.~\cite[p.~209]{curtis81}. The idempotent
$e_\chi$ is hermitian, since $\ordrho$ is unitary, hence an
orthogonal projection operator. That $e_\chi$ projects onto $Q$ is a
well-known fact, cf. Theorem~8 in~\cite[p.~21]{serre77}.  The
dimension of the module $W$ is $\chi(1)$, whence
$\dim_\C(Q)=m\chi(1)$.

\item[\textit{Step 2.}] 
Let $g,h\in E$. The characters of $gW$ and $hW$ are
$\psi(x)=\chi(gxg^{-1})$ and $\varphi(x)=\chi(hxh^{-1})$ respectively,
cf.~Chap.~V, \S17, Theorem~17.3 c) in~\cite{huppert83}.  Suppose that
$g$ and $h$ are not in the same coset of $T(W)$ in $E$. Then $\psi$
and $\varphi$ are different irreducible characters. Thus the
idempotents $e_\psi$ and $e_\varphi$ satisfy $e_\chi
e_{\psi}=0=e_{\psi} e_\chi$, hence project on orthogonal subspaces. We
have $\image(e_\psi)=gQ$, $\image(e_\varphi)=hQ$, and thus, in
particular, $gQ\,\bot\, hQ$.

\item[\textit{Step 3.}] It remains to show the error correcting properties of
$Q$. Recall that an error $w$ can be detected if and only if $e_\chi
\ordrho(w) e_\chi$ is a scalar multiple of $e_\chi$, cf.~\cite{knill96b}.
The code $Q$ is $\Sigma$-correcting if and only if $Q$ is able to
detect all errors in $\{ e_1^{-1}e_2\,|\, e_1, e_2\in \Sigma\}$,
cf.~\cite{BDSW96,KnLa97}. Hence it remains to show that an error $w$
can be detected if and only if $w\not\in T(W)-Z(W)$.

\begin{enumerate}
\item An error $w\in Z(W)$ can be detected, since, by definition, there
exists a scalar $\lambda \in \C$ such that $e_\chi\ordrho(w)e_\chi=\lambda
e_\chi$. 

\item An error $w\in E - T(W)$ can be detected, since Step 2 shows
that $e_\chi\ordrho(w)e_\chi=0$ holds.

\item An error $w\in T(W)-Z(W)$ cannot be detected. Indeed, $\ordrho(w)$
maps $Q$ into itself, since $w\in T(W)$. However, $e_\chi
\ordrho(w)e_\chi$ cannot be a multiple of $e_\chi$, since this would
imply that $w$ is an element of $Z(W)$. 
\end{enumerate}
This proves the claim.~\qed
\end{enumerate}

\noindent The error correcting properties of a Clifford code $Q$ are fully
determined by the inertia group $T(W)$ and the group $Z(W)$. 
It is often more convenient to use characters rather than
modules to compute these groups. The inertia group 
$T(W)$ coincides with the inertia group $T(\chi)$ of the character $\chi$ in $G$:
$$ T(W)=T(\chi)=\{ g\in G\,|\, \chi(gxg^{-1})=\chi(x)\;\mbox{for
all}\; x \in N\}.$$ The group $Z(W)$ can also be determined by a
character. Clifford theory shows that $Q$ is an irreducible $\C
T$-module, where $T=T(W)$. Denote by $\vartheta$ the irreducible character of $T$
afforded by~$Q$. Then $Z(W)$ is determined by the values of the
character~$\vartheta$: 
$$ Z(W)=Z(\vartheta)=\{ g\in T\,|\,\vartheta(1)=|\vartheta(g)|\,\}.$$

\section{Characters}
We have seen that the inertia group of $\chi$ determines the error
correcting properties of the quantum code $Q$. We show in this section
how the inertia groups can be calculated for abstract error groups
with abelian index groups.

Let us first recall a few standard notations from group theory.  If $E$
is a finite group, then $E'$ denotes the commutator subgroup, 
$$ E' = \langle [g,h]=g^{-1}h^{-1}gh\,|\, g,h\in E\rangle.$$
The center $Z(E)$ of $E$ is given by the group 
$$ Z(E)=\{ z\in E\,|\, zg=gz\;\mbox{for all}\; g\in E\}.$$

An abstract error group $E$ has an abelian index group $G\cong E/Z(E)$
if and only if its commutator subgroup $E'$ is contained in $Z(E)$.
For that reason, it is of interest to study the inertia groups in such
groups $E$.  We will see that the inertia group of a character $\chi$
of $N$ defining a Clifford code is simply given by the centralizer of
$Z(N)$ in~$E$.

Let $G$ be a finite group. We denote by
$\Irr(G)$ the set of irreducible characters of $G$. We say that a
character $\chi\in\Irr(G)$ is faithful on $H\subseteq G$ if and only
if the intersection of $H$ with the kernel of $\chi$ is trivial
$$ H \cap \ker(\chi) = H \cap \{ g\in
G\,|\, \chi(1)=\chi(g)\} =\{1\}.$$
We need to establish a few simple properties of characters. 
We will see that a character defining a Clifford code will satisfy the
assumption of the following lemma, which gives some information about
character values. 
\begin{lemma}\label{charvalue}
Let $E$ be a finite group, $N\trianglelefteq E$. Let $\chi$ be an
irreducible character of $N$ that is faithful on $Z=Z(E)\cap N$. If $z\in
Z$, $z\not= 1$, and $n\in N$, then $\chi(zn)=\omega\chi(n)$ for some
$\omega\not=1$. 
\end{lemma}
\proof Denote by $\ordrho$ a representation affording $\chi$. Since
$\ordrho$ is irreducible, 
$\ordrho(z)$ is a scalar multiple of the identity matrix $I$ for all
$z\in Z$ by Schur's lemma. If $z\not= 1$, then $\ordrho(z)=\omega I$ with
$\omega\not=1$, since $\chi$ is faithful on~$Z$.
Hence
$\chi(zn)=\trace(\ordrho(zn))=\trace(\omega\ordrho(n))=\omega\trace\ordrho(n)=\omega\chi(n)$
as claimed.~\qed 

In the next step we want to show that the character $\chi$ defining a
Clifford code is indeed faithful on the central elements of $E$
contained in $N$. We exploit the fact that $\chi$ is a constituent of
a faithful character $\phi\in \Irr(E)$ of the abstract error group 
satisfying $\phi(1)^2=(E\!:\!Z(E))$. 
Recall that a scalar product of two characters $\chi,\vartheta\in
\Irr(N)$ is defined by 
$$ \langle\chi,\vartheta\rangle=\frac{1}{|N|} \sum_{n\in N} \chi(n)\vartheta(n^{-1}).$$
This allows to define the set of irreducible components of the restriction
of $\phi\in \Irr(E)$ to $N$ by 
$$ \IrrP(N)=\{ \chi\in \Irr(N)\,|\, \langle \chi, \phi\!\downarrow\!N\rangle\not=0\},$$
where $\phi\!\downarrow\! N$ denotes the restriction of $\phi$ to $N$. 
Using this notation, we can now formulate
\begin{lemma}\label{faithchar}
Let $E$ be a finite group, $N\trianglelefteq E$, $\phi\in \Irr(E)$, 
and $\chi\in
\IrrP(N)$. If $\phi$ is faithful on $Z(E)$, then $\chi$ is
faithful on $Z=Z(E)\cap N$. 
\end{lemma}
\proof By Clifford's theorem, the restriction of $\phi$ to $N$ can be
expressed as a sum of characters 
$\chi^g(x)=\chi(gxg^{-1})$ conjugated to $\chi$: 
$$ (\phi\downarrow N)(x)=m\sum_{g\in R} \chi^g(x),$$
for some subset $R$ of $E$. 
The conjugated characters satisfy
$\chi^g(z)=\chi(gzg^{-1})=\chi(z)$ for all central elements $z\in Z$. 
Hence
$$ (\phi\downarrow N)(z)=|R|m\chi(z)$$
for all $z\in Z$, which proves the claim.~\qed

Recall that the support $\supp(\chi)$ of a function $\chi\colon E\rightarrow \C$ is given by the set $ \supp(\chi)=\{ g\in E\,|\, \chi(g)\neq 0\}.$
We use our knowledge of character values to determine the support
of the character $\chi$: 
\begin{lemma}\label{suppchar}
Let $E$ be a finite group satisfying
$E'\subseteq Z(E)$, and $N\trianglelefteq E$. 
If $\chi\in \Irr(N)$ is faithful on
$Z= N\cap Z(E)$, then\/ $\supp(\chi)=Z(N)$. 
\end{lemma}
\proof Let $n\in \supp(\chi)$. Seeking a contradiction, we assume that
$n\not\in Z(N)$. Since $E'\subseteq Z(E)$, this means that there
exists an element $g\in N$ such that $gng^{-1}=zn$ for some $z\in
Z(E)$, $z\not=1$. Note that $zn$, hence $z$, is an element of $N$ since
$N$ is a normal subgroup of $E$. Thus,
$$\chi(n)=\chi(gng^{-1})=\chi(zn)=\omega\chi(n)$$
with $\omega\not=1$, by Lemma~\ref{charvalue}. 
This contradicts the fact that $\chi(n)\not=0$, hence
$\supp(\chi)=Z(N)$ as claimed.~\qed

Recall that the centralizer $C_E(H)$ of a subgroup $H$ in $E$ is given by the group  
$$ C_E(H)= \{ g\in E\,|\, ghg^{-1}=h\;\mbox{for all}\; h\in H\}. $$
Using this notation, we are able to explicitly determine the inertia subgroup $T(\chi)$: 

\begin{lemma}[``Tacheles" Lemma]\label{keylemma}
Let $E$ be a finite group satisfying $E'\subseteq Z(E)$, and
$N\trianglelefteq E$. 
Let $\phi\in \Irr(E)$ be faithful on $Z(E)$, and $\chi\in
\IrrP(N)$. Then the inertia group of $\chi$ in $E$ is given by 
$T(\chi)=C_E(Z(N))$. 
\end{lemma}
\proof 
The character $\chi$ is faithful on $Z(E)\cap N$ by Lemma~\ref{faithchar}. 
Thus $\supp(\chi)=Z(N)$ by Lemma~\ref{suppchar}. It follows that
$C_E(Z(N))\le T(\chi)$. Conversely, suppose that $g\not\in
C_E(Z(N))$. We want to show that $g$ cannot be an element of the
inertia group. 
Since $E'\subseteq Z(E)$, the condition  $g\not\in
C_E(Z(N))$ implies that there
exists an element $n\in Z(N)$ such that $gng^{-1}=zn$ for some $z\in
Z(E)$, $z\not=1$.
Since $N$ is a normal subgroup of $E$, we also obtain that $zn\in N$.
Together with $n\in N$ this shows that $z\in N$. 
By Lemma~\ref{charvalue},
$\chi^g(n)=\chi(gng^{-1})=\chi(zn)=\omega\chi(n)$ with
$\omega\not=1$. Since $n\in Z(N)\subseteq \supp(\chi)$,
$\chi(n)\not=0$, whence $g\not\in T(\chi)$.~\qed

\section{Abelian Index Groups}
Suppose that we fix a normal subgroup 
$N$ of an abstract error group $E$ and define a Clifford code $Q$
using a character $\chi\in \IrrP(N)$. If the index group of $E$ is
abelian, then the next theorem shows that $Q$ could have been derived
from an abelian group, namely from the center $Z(N)$ of $N$. 

\begin{theorem}\label{clifford}
Let $E$ be an abstract error group with abelian index group. 
Let $N$ be a normal subgroup of $E$. Suppose that $Q$ is a Clifford
code with respect to $N$, then $Q$ is also a Clifford code with
respect to $Z(N)$.
\end{theorem}
\proof We divide the proof into several steps.  
\begin{enumerate}
\item[\textit{Step 1.}] 
The Clifford code $Q$ is defined by the following data. 
There exists a faithful irreducible character $\phi$ of $E$ that
corresponds to a unitary representation $\ordrho$ of degree
$(E\!:\!Z(E))^{1/2}$ and $\chi\in \IrrP(N)$ such that 
$$ e_\chi=\frac{\chi(1)}{|N|}\sum_{n\in N} \chi(n^{-1})\ordrho(n)$$ 
is an orthogonal projector onto $Q$. 
\goodbreak

\item[\textit{Step 2.}]  Recall that $E$ satisfies $E'\subseteq Z(E)$,
since the index group $E/Z(E)$ is abelian. We want to show that
$N\trianglelefteq E$ implies that $Z(N)\trianglelefteq E$. Indeed,
take $n\in Z(N)$ and $g\in E$. We have $gng^{-1}=zn$ for some $z\in
Z(E)$, since $E'\subseteq Z(E)$. Now $zn\in N$, since
$N\trianglelefteq E$, and thus $z\in N$.  On the other hand, an
element $z\in Z(E)\cap N$ is an element of $Z(N)$. This shows that all
conjugates of an element $n\in Z(N)$ are again elements of $Z(N)$,
whence $Z(N)\trianglelefteq E$.
\goodbreak

\item[\textit{Step 3.}]  The restriction of $\chi$ to the center
$Z=Z(N)$ is given by $(\chi\!\downarrow\!Z)(x)=\chi(1)\,\varphi(x)$
for some irreducible character $\varphi$ of~$Z$, cf.~Prop. 6.3.5
in~\cite{grove97}.  We claim that
$$ e_\varphi = \frac{1}{|Z|} \sum_{z\in Z} \varphi(z^{-1})\ordrho(z)$$
is also an orthogonal projector onto $Q$.
It is clear that $\dim_\C \image(e_\chi)= \dim_\C
\image(e_\varphi)$, since the ``Tacheles" Lemma shows that the inertia 
groups of $\chi$ and $\varphi$ are given by 
$T(\chi)=C_E(Z)=T(\varphi)$. 
Thus, it suffices to show that the dimension of $\image(e_\varphi
e_\chi)$ is not smaller than the dimension of
$\image(e_\chi)$.
\goodbreak

\item[\textit{Step 4.}]
Recall that $\phi(g)=\trace \ordrho(g)$ is zero for all $g\in E$
not in the center $Z(E)$. Moreover, $(\phi\!\downarrow\!Z)(z)=
\phi(1)\varphi(z)$ holds for all $z\in Y=Z(E)\cap N$,
cf.~Lemma~\ref{faithchar}.  Keeping this in mind, it is easy to
calculate the dimension of $\image(e_\chi)$ by
$$ \dim_\C\image(e_\chi)=\trace e_\chi =  \frac{|Y|\,\phi(1) \chi(1)^2 }{|N|}.$$
On the other hand, we find that 
$$ 
\begin{array}{l}
\dim_\C\image(e_\varphi e_\chi)=\trace(e_\varphi e_\chi)\\[1ex]
\qquad= \displaystyle\frac{\chi(1)}{|N|\,|Z|}\sum_{n\in N,z\in Z\atop nz\in Y}
\varphi(z^{-1})\chi(n^{-1})\trace\ordrho(nz).
\end{array}
$$
Since $\chi(z)=\chi(1)\varphi(z)$ holds for $z\in Z$, and the
conditions $z\in Z$ and $nz\in Y$
imply that $n\in Z$, we can further simplify this
expression to
$$
\begin{array}{lcl}
\trace(e_\varphi e_\chi)&=&\displaystyle\frac{\chi(1)^2}{|N|\,|Z|}\sum_{n, z\in Z\atop nz\in Y}
\varphi(z^{-1}n^{-1})\trace\ordrho(nz)\\[1ex]
&=& \displaystyle
\frac{|Y||Z|\phi(1)\chi(1)^2 }{|N||Z|}.
\end{array}
$$
This shows that $\dim_\C\image(e_\chi)=\dim_\C\image(e_\varphi
e_\chi)$, whence 
$e_\chi$ and $e_\varphi$ project both onto
$Q$.~\qed 
\end{enumerate}

\section{Conclusions}
We have shown some basic properties of Clifford codes, which are a
natural generalization of stabilizer codes. The main result of this
note shows that there is no loss in assuming that the normal subgroup
defining a Clifford code is abelian provided that the abstract error
group is a nilpotent group of class at most~2.  An analogue of
Theorem~\ref{clifford} does not hold for general index groups. In
fact, we have recently shown that there exist Clifford codes which are
not stabilizer codes~\cite{klappenecker02}.  This result indicates
that abstract error groups with nonabelian index groups might provide
a new angle to the theory of quantum error correcting codes. Moreover,
this shows that the theory of Clifford codes~-- after all~-- extends the
concept of stabilizer codes.


\begin{thebibliography}{10}

\bibitem{ashikhmin00}
A.~Ashikhmin and E.~Knill,
\newblock ``Nonbinary quantum stabilizer codes,''
\newblock {\em IEEE Trans. Inform. Theory}, vol. 47, no. 7, pp. 3065--3072,
  2001.

\bibitem{calderbank98}
A.R. Calderbank, E.M. Rains, P.W. Shor, and N.J.A. Sloane,
\newblock ``Quantum error correction via codes over {GF}(4),''
\newblock {\em IEEE Trans. Inform. Theory}, vol. 44, pp. 1369--1387, 1998.

\bibitem{gottesman96}
D.~Gottesman,
\newblock ``Class of quantum error-correcting codes saturating the quantum
  {H}amming bound,''
\newblock {\em Phys. Rev. A}, vol. 54, pp. 1862--1868, 1996.

\bibitem{rains99}
E.M. Rains,
\newblock ``Nonbinary quantum codes,''
\newblock {\em IEEE Trans. Inform. Theory}, vol. 45, pp. 1827--1832, 1999.

\bibitem{klappeneckerPartI}
A.~Klappenecker and M.~R{\"o}tteler,
\newblock ``Beyond stabilizer codes {I}: Nice error bases,''
\newblock this volume.

\bibitem{knill96b}
E.~Knill,
\newblock ``Group representations, error bases and quantum codes,''
\newblock Los Alamos National Laboratory Report LAUR-96-2807, 1996.

\bibitem{huppert83}
B.~Huppert,
\newblock {\em Endliche Gruppen I},
\newblock Springer-Verlag, Berlin, 2nd edition, 1983.

\bibitem{curtis81}
C.W. Curtis and I.~Reiner,
\newblock {\em Methods of Representation Theory}, vol.~{I},
\newblock John Wiley \& Sons, New York, 1981.

\bibitem{serre77}
J.-P. Serre,
\newblock {\em Linear Representation of Finite Groups},
\newblock Springer-Verlag, New York, 1977.

\bibitem{BDSW96}
C.H. Bennett, D.P. DiVincenzo, J.A. Smolin, and W.K. Wootters,
\newblock ``{Mixed State Entanglement and Quantum Error Correction},''
\newblock {\em Physical Review~A}, vol. 54, no. 5, pp. 3824--3851, 1996.

\bibitem{KnLa97}
E.~Knill and R.~Laflamme,
\newblock ``{A theory of quantum error--correcting codes},''
\newblock {\em Physical Review~A}, vol. 55, no. 2, pp. 900--911, 1997.

\bibitem{grove97}
L.C. Grove,
\newblock {\em Groups and Characters},
\newblock John Wiley, New York, 1997.

\bibitem{klappenecker02}
A.~Klappenecker and M.~R{\"o}tteler,
\newblock ``Clifford codes,''
\newblock in {\em The Mathematics of Quantum Computing}, R.~Brylinski and
  G.~Chen, Eds. 2002, CRC-Press,
\newblock to appear.

\end{thebibliography}

\end{document}